\documentclass[12pt]{iopart}
\usepackage{iopams}
\usepackage{graphicx}

\usepackage{color}

\begin{document}

\definecolor{Red}{rgb}{1,0,0}

\newcommand{\CB}{{\cal B}}
\newcommand{\CC}{{\cal C}}
\newcommand{\CD}{{\cal D}}
\newcommand{\CE}{{\cal E}}
\newcommand{\CF}{{\cal F}}
\newcommand{\CH}{{\cal H}}
\newcommand{\CJ}{{\cal J}}
\newcommand{\CK}{{\cal K}}
\newcommand{\CQ}{{\cal Q}}
\newcommand{\CR}{{\cal R}}
\newcommand{\CU}{{\cal U}}
\newcommand{\CW}{{\cal W}}
\newcommand{\average}[1]{\left\langle #1 \right\rangle_\CD}
\newcommand{\laverage}[1]{\left\langle #1 \right\rangle_{\CD_{\rm \bf i}}}
\newcommand{\baverage}[1]{\left\langle #1 \right\rangle_{\CD_R}}
\newcommand{\averageM}[1]{\left\langle #1 \right\rangle _{{\cal M}}}
\newcommand{\averageE}[1]{\left\langle #1 \right\rangle _{{\cal E}}}
\newcommand{\averageA}[1]{\left\langle #1\right\rangle _{\CA}}
\newcommand{\averageF}[1]{\left\langle #1\right\rangle _{\CF}}
\newcommand{\averageDN}[1]{\left\langle #1\right\rangle _{\CD_{0}}}
\newcommand{\averageFi}[1]{\left\langle #1\right\rangle _{\CF_{\ell}}}
\newcommand{\initial}[1]{{#1_{\rm \bf i}}}
\newcommand{\inI}{{\bf I}}
\newcommand{\inII}{{\bf II}}
\newcommand{\inIII}{{\bf III}}


\article[Effective inhomogeneous inflation]{FAST TRACK COMMUNICATION}{Effective inhomogeneous inflation: \\ curvature inhomogeneities of the Einstein vacuum}
\author{Thomas Buchert$^{1}$ and Nathaniel Obadia$^{2}$}
\address{$^{1}$Universit\'e Lyon~1, Centre de Recherche Astrophysique de Lyon,
\\9 Avenue Charles Andr\'e, F--69230 Saint--Genis--Laval,
France}

\address{$^{2}$\'Ecole~Normale~Sup\'erieure de Lyon, Centre de Recherche
Astrophysique de Lyon,\\
46 All\'ee d'Italie, F--69364 Lyon Cedex 07, France\\ \bigskip
Emails: buchert@obs.univ--lyon.fr and nathaniel.obadia@ens--lyon.fr}
%


\begin{abstract}
We consider spatially averaged inhomogeneous universe models and argue that, already in the absence of sources, 
an effective scalar field arises through foliating and spatially averaging inhomogeneous geometrical curvature invariants of the Einstein vacuum.
This scalar field (the `morphon') acts as an inflaton, if we prescribe a potential of some generic form. 
We show that, for any initially negative average spatial curvature, the morphon is driven through an inflationary phase and leads -- on average -- to a spatially flat, homogeneous and isotropic universe model, providing initial conditions for pre--heating and, by the same mechanism, a possibly natural self--exit.
\end{abstract}

\section{Introduction} 

For about $30$ years now, inflation \cite{Starobinsky:1980te}
is the main paradigm 
used to explain the various caveats of the Hot Big Bang scenario. 
In particular, the flatness, smoothness and horizon issues \cite{Bassett:2005xm} are
often advocated to be solved by a long enough
period of de Sitter--like expansion, provoked by
of a slow--rolling scalar (multi--)field that dominates over the other components
of the energy budget of the Universe \cite{Liddle:1994dx}.  
The success of the concept of inflation stems not only from its simplicity but also 
from the consequences this framework bears, 
{\it i.e.} conservative predictions concerning scale invariance, 
the tensor--to--scalar ratio, Gaussianity of fluctuations, etc. 
However, as many paradigms praised for their consequences, 
inflation lacks a physical cause. 
Indeed, despite several attempts to justify the global predominance of 
a yet unobserved  fundamental scalar field,
the existence of the inflaton remains a conundrum to theorists. 

Confronted with this issue, physicists might roughly choose between three distinct philosophies. 
Either a non--standard part of the particle physics spectrum,
for instance, the Higgs boson \cite{Barbon:2009ya}, or a pseudo Nambu--Goldstone boson \cite{Dolgov:1994zq},
could be candidates for the fundamental field whose potential part dominated the Early Universe. 
Or, a known part of the physical fields could emulate an effective scalar field 
via a known process like boson or fermion condensation \cite{Parker:1993ya,Shankaranarayanan:2009sz}.
Or, a neglected part in the underlying theories, due to their special nature or just due to idealizing
assumptions, prevented us so far to take into account physical effects that could
lead to an inflationary era. 
The idea we present here lies within this last line of thought
by applying to inflation what has been proposed previously
for a conservative explanation of the dark energy problem through 
inhomogeneities \cite{Buchert:2007ik}.
In concrete terms we shall provide a physical cause to inflation 
by identifying the inflaton with an 
effective classical field that describes the averaged bulk effect 
on the dynamics of inhomogeneities (also dubbed backreaction) 
prior to, during and eventually after the inflationary era. 

It is important to stress that the model expounded here
pertains to the {\it dynamics} of backreaction 
-- described by a conservative set of equations 
and a minimal number of physical assumptions --
and not to a suitably chosen set of initial conditions. 
Furthermore, as a natural starting point,
the inhomogeneities we shall describe hereby are due 
to geometry and {\it a priori} not related
to the baryonic/dark matter, dark energy or radiation content of the Universe \cite{Lifshitz:1963ps}; 
we shall confine ourselves to inhomogeneities in the gravitational field only.
In other words, this model will demonstrate in simple terms
how inflation can naturally emerge out of {\it vacuum inhomogeneities}.
This last term does not refer to quantum vacuum fluctuations
on a  FLRW background (as practically always
used in the cosmological context),
but to the average of classical fluctuations of the gravitational field only. 

\section{Inhomogeneous universe models}

Einstein's equations are assumed to hold 
for a four--dimensional tube of the Early Universe, featuring the
$4-$Ricci tensor 
${R}_{\mu\nu} = 0$\footnote{Greek indices run through $0,1,2,3$, while latin indices run through $1,2,3$, and we set $c=8\pi G=1$.}.
We suggest further to foliate this tube into three--dimensional space--like
hypersurfaces according to the ADM formalism.
We here choose a comoving--synchronous line--element 
(in forthcoming papers the embedding issue will be thoroughly addressed),
$ds^{2}=-dt^{2}+g_{ij}dX^{i}dX^{j}$, where the proper time $t$ labels the
hypersurfaces and $X^i$ are Gaussian normal coordinates;
$g_{ij}$ are the components of the full inhomogeneous $3-$metric of the
hypersurfaces of constant proper time. In this metric the components of the
$4-$Ricci tensor can be expressed geometrically through the three--dimensional
extrinsic curvature $K_{ij}$ of the embedding into spacetime, and
instrinsic curvature ${\cal R}_{ij}$ of the hypersurfaces at constant $t$ with
the following well--known Gau{\ss}--Codazzi--Mainardi relations:
\begin{equation}
\fl\quad
R_{00} = {\dot K} - K^{ij}\,K_{ij}\;\;,\;\; 
R_{0i}= K_{|i} - K^k_{\;i||k}\;\;,\;\;
R_{ij}={\cal R}_{ij} - {\dot K}_{ij}-2K_{ik}K^{k}_{\;j}+ K K_{ij}\;\;,
\end{equation}
where a vertical slash denotes partial spatial derivative with respect to $X^i$, 
a double vertical slash covariant spatial differentiation with respect to the $3-$metric,
and an overdot the covariant time derivative. 
We end up with the following set of equations
(supplemented by the defining equation for $K_{ij}$):
\begin{equation}
\label{admvacuum}
\fl
{\dot g}_{ij} = -2\,K_{ij}\:,\: -{\dot K} +  K^{ij}\,K_{ij} \,=\, 0\;\;;\;\;
K_{|i} - K^k_{\;i||k}\,=\,0\;\;;\;\;
 - {\dot K}_{ij}+ K K_{ij} -2K_{ik}K^k_{\;j}\,=\,{\cal R}_{ij}\,.
\end{equation}
The second equation is the vacuum version of Raychaudhuri's equation, the third
set of equations are the momentum constraints. Forming the trace of the last
equation and inserting it into Raychaudhuri's equation, we obtain the vacuum
version of the Hamiltonian constraint:
$
{\cal R} + K^2 - K^i_{\;j}\,K^j_{\;i} \,=\,0 
$,
where ${\cal R}= {\cal R}^k_{\;k}$ is the scalar curvature with respect to the
$3-$metric, and $K := K^k_{\;k} =-\Theta $ can be interpreted as (minus) the local
expansion rate of the hypersurfaces.
We further propose to spatially average the scalar parts of the above equations, defined
for any scalar function $\Psi (t,X^i)$ as
\begin{equation}
\label{average}
\langle \Psi (t, X^i)\rangle_{\cal D}: =
\frac{1}{V_{\cal D}}\int_{\cal D}  \;\Psi (t, X^i) \;\sqrt{\det(g_{ij})} d^3 X\ ,
\end{equation}
where the volume of an arbitrary compact domain is
$V_{\cal D}(t) : = \int_{\cal D} \sqrt{\det(g_{ij})} d^3 X$. 
Defining a volume scale factor by
$a_{\CD}\left(t\right):=\left(V_{\cal D}(t)/V_{\cal D}(t_{i})\right)^{1/3}$,
and averaging Raychaudhuri's equation and the Hamiltonian constraint 
using the non--commutativity relation (true for any scalar $S$),
$\langle S\rangle\dot{}_\CD - \langle{\dot S}\rangle_\CD = \average{\Theta \: S} - \average{\Theta} \average{S}$,
we obtain the
following well--known equations \cite{Buchert:1999er}:
\begin{equation}
\frac{\ddot{a}_{\CD}}{a_{\CD}}  =  \frac{\CQ_{\CD}}{3}
\;\;\;;\;\;\;
H_{\CD}^{2}  = - \frac{ k_\initial\CD}{a_\CD^2} -\frac{1}{6} \left( \CW_\CD  + \CQ_{\CD}\right) \;,
\label{averagedequations}
\end{equation}
where $H_{\CD}$ denotes the (domain--dependent) Hubble rate 
$H_{\CD}=\dot{a}_{\CD} / a_{\CD}=-1/3\average{K}$, and $k_\initial\CD$ a (domain--dependent) constant of integration. The kinematical backreaction $\CQ_{\CD}$ is
composed of extrinsic curvature invariants, while $\CW_\CD$ is an intrinsic
curvature invariant and describes the deviation from constant--curvature,
\begin{equation}
\CQ_{\CD}:= \average{K^2 - K^i_{\;j}K^j_{\;i}} - \frac{2}{3}\average{K}^2 \;\;;\;\;\CW_{\CD}: = \average{\CR} - \frac{6 k_\initial\CD}{a_\CD^2}\;.
\label{eq:Def-Q}
\end{equation}
For a homogeneous domain the above backreaction terms, being covariantly defined with respect to a given spatial embedding, are zero. 
Therefore they encode the departure from spatial homogeneity. The integrability condition connecting the two Eqs.(\ref{averagedequations}), 
reads:
\begin{equation}
a_{\CD}^{-2} (a_{\CD}^{2}\CW_\CD \dot{)}\,=\,-a_{\CD}^{-6} (a_{\CD}^{6}\CQ_{\CD}\dot{)}\;,
\label{eq:integrability}
\end{equation}
expressing a combined conservation law for intrinsic curvature and extrinsic fluctuations.

\section{Effective scalar field: the morphon} 

We rewrite the set of spatially averaged equations (\ref{averagedequations},\ref{eq:integrability}) and cast it into
a Friedmannian form with a purely geometrical effective energy--momentum
tensor \cite{Buchert:2001sa,Buchert:2005kj}:
\begin{equation}
\fl\quad\quad
3\frac{{\ddot{a}}_{\CD}}{a_{\CD}}  =  -\frac{\varrho_{\rm eff}^{\CD}+3{p}_{\rm eff}^{\CD}}{2}\;\;;\;\;
3H_{\CD}^{2}+ \frac{3 k_\initial\CD}{a_\CD^2}= \varrho_{\rm eff}^{\CD} \;\;;\;\;
{\dot{\varrho}}_{\rm eff}^{\CD}+3H_{\CD}\left(\varrho_{\rm eff}^{\CD}+{p}_{\rm eff}^{\CD}\right) =  0 \;,
\end{equation}
where the effective densities are ``morphed'' by a (minimally coupled) scalar field, 
the {\em morphon}  $\Phi_\CD$, thoroughly discussed in \cite{Buchert:2006ya} and here defined through:
\begin{equation}
\fl\quad\quad
\varrho_{{\rm eff}}^{{\CD}}  :=  -\frac{{\CQ}_{{\CD}}+{\CW}_{{\CD}}}{2}
= \xi \frac{1}{2}{{\dot\Phi}_\CD}^2 + U_\CD \quad;\quad
{p}_{{\rm eff}}^{{\CD}}  :=  -\frac{3{\CQ}_{{\CD}}-{\CW}_{{\CD}}}{6}
=\xi \frac{1}{2}{{\dot\Phi}_\CD}^2 - U_\CD \;,
\end{equation}
where $\xi=+/-1$ for a standard/phantom scalar field,
and $U_\CD = U_\CD (\Phi_\CD)$ is the self--interaction potential. 
From the above equations we obtain the following correspondence:
\begin{equation}
\label{correspondence1}
{\CQ}_\CD \;=\; U_\CD- \xi {\dot\Phi}^2_\CD  \;\;\;;\;\;\;
\CW_{\CD}= -3 U_\CD\;\;.
\end{equation}
We appreciate that the deviation 
of the averaged scalar curvature from a constant--curvature model $\CW_{\CD}$
is directly proportional to the potential energy; 
an expanding (or contracting) classical vacuum with on average negative scalar curvature deviation (a positive potential $U_{\CD}$) 
can be attributed to a negative potential energy of a morphon field. The homogeneous case, $\CQ_\CD = 0$,
corresponds to a virial equilibrium of the scalar field energies.
Inserting (\ref{correspondence1}) into the integrability condition (\ref{eq:integrability}) 
implies that $\Phi_\CD$, for ${\dot\Phi}_\CD \ne
0$, obeys the (scale--dependent) Klein--Gordon equation:
\begin{equation}
\label{eq:kleingordon}
{\ddot\Phi}_\CD + 3 H_{\cal D}{\dot\Phi}_\CD +
\xi\frac{\partial}{\partial \Phi_\CD}U_\CD \;=\;0\;\;.
\end{equation}
Averaged universe models obeying the previous set of equations follow, thus, 
Friedmannian kinematics with no fundamental sources, 
but with an effective scalar field perfect fluid that reflects the shape of spatial hypersurfaces 
and their embedding into spacetime.
Given a choice of $\xi$ and of the potential, the evolution of these models is fixed (the governing equations are closed). 

In order to model inflation we impose the constraints
$(U_\CD > 0,\xi=+1)\Rightarrow(\CW_\CD<0,\: \CW_\CD + 3 \CQ_\CD < 0)$:
the morphon is canonical and its potential competes with its kinetic energy.
These conditions still allow for any sign of the kinematical backreaction term; 
rewriting the extrinsic curvature invariants in (\ref{eq:Def-Q}) 
in terms of the local expansion and shear \cite{Buchert:2007ik}, 
we get that geometries dominated by their expansion fluctuations have $\CQ_\CD>0\Leftrightarrow U_\CD >\dot\Phi_\CD^2$ (hence $\ddot a_\CD>0$, see (\ref{averagedequations})), 
those dominated by their shear fluctuations, $\CQ_\CD<0$; and homogeneous spacetimes obey $\CQ_\CD=0$ on all scales.

\section{A morphonic inflaton} 

We offer the idea that the morphon can formally play the role of the inflaton, 
and that (unavoidable) curvature inhomogeneities occurring at one point 
of the Universe's history
could be the actual cause of a de Sitter--like era prior to the baryon nucleosynthesis era,
provided they are governed by an appropriate potential.
In the average quantities dictionary, inflation can be read off in terms of the effective
first slow--roll parameter 
\begin{equation} 
\ddot{a}_{\CD}> 0 \; \Leftrightarrow \; \epsilon := - \dot H_\CD/H_\CD^2 < 1 \ .
\end{equation}
Some remarks are already necessary at this stage
in order to assess our paradigm and to
distinguish it from the usual ``FLRW$+$fundamental sources" inflationary picture. 
First, since Eqs.~(\ref{averagedequations}) only functionally depend on a metric that needs not to be specified, 
requiring $\ddot{a}_{\CD}> 0$ does not imply that the comoving Hubble distance $1/a H$ decreases;
however, such is the essence of the fitting problem
to find the best FLRW fitting model to a lumpy universe \cite{Ellis:1987zz,ellisbuchert}.
Secondly, a fair solution of the flatness and smoothness problems
would necessitate to implement averaging on the light cone, or would require an explicit inhomogeneous metric.
Lacking the latter in the present context, we claim that
realizing $\epsilon < 1$ during a sufficient number of e--folds is the best condition to address the dynamics of inflation.
Third, so as to legitimate the neglect of any averaged matter, we refer to \cite{phasespace}, where it is shown with the same class of assumptions 
(a flat enough curvature invariant $\CW_\CD$, see (\ref{GLpotential}) below) that the corresponding cosmological parameter $\Omega_m^\CD=\average{\rho}/3H_\CD^2$ dives under a percent within a few $H_\CD^{-1} (t_i)$ times.
Therefore a large enough initial ``Hubble parameter" $H_\CD (t_i)$ consigns the matter content to oblivion. 
 
A suitable potential $U_\CD$ has to be chosen 
in order to identify the morphon as a trustworthy inflaton.
This identification implies that the cause (the inhomogeneities)
and the consequence (the smoothness) 
of the present model are at odds.
Therefore, a meaningful potential needs not only to describe how inhomogeneities can induce inflation,
but also how they can disappear throughout this process. 
Translated to the standard inflationary paradigm,
this condition can be easily implemented by any potential that possesses 
a minimum downwards which the inflaton rolls.
One of the simplest examples, which has been extensively studied in the context
of chaotic inflation \cite{Linde:1983gd}, is a potential of the Ginzburg--Landau form:  
\begin{equation}
\label{GLpotential}
U_\CD^{GL} = U_0 \left(\Phi_\CD^2 - \Phi_0^2 \right)^2/\Phi_0^4 \;.
\end{equation}
This quartic potential can also be related to a fundamental Higgs field. 
However, even if such a scalar field is fundamental, 
there is the possibility that no Higgs particle is involved -- as in our case 
-- {\it e.g.} \cite{Dehnen:1992jc}.
Contrary to standard inflation one must here raise the issue of the ``reality'' of such a potential, since
we postulate that  $U_\CD$ is actually given by the averaged scalar curvature (\ref{correspondence1}).
More precisely, one should question the assumption of a slow--roll period, 
which corresponds to a nearly constant value of the averaged curvature. 
At this stage we cannot provide a proof for this possibility, 
{\it e.g.} through a suitable exact solution or through a well--defined approximation for the local evolution of curvature, 
but we can summarize plausible hints that the actual physical properties of an intrinsic curvature distribution could comply with the expected:
first, there are indications from perturbation theory that the perturbative expansion is led by a term $\CW_\CD \propto a_\CD^{-1}$, 
when putting the Friedmannian zero--order curvature to zero (so that the scenario is exclusively governed by backreaction, i.e. the curvature--deviation from a flat model), 
and this perturbative expansion can be extended up to including even a constant term \cite{Li:2007ci}; 
secondly, a concrete modeling of the non--perturbative aspects of the curvature distribution in a multiscale analysis reveals that, 
even if subdomains are evolving according to the term $\CW_\CD \propto a_\CD^{-1}$, 
the variance between subdomains can lead to a de Sitter--like behavior on large scales \cite{multiscale}.
These results, though not being a proof of the actual choice we made (say, we employ  (\ref{GLpotential}) as an illustrative example),
open the possibility that there exist some configurations of the geometrical inhomogeneities 
that enable a slowly varying curvature, that is a sufficiently flat potential. Obviously, we here are in the same situation as for an explanation of the 
dark energy phenomenon through inhomogeneities, an ongoing research field that is qualitatively well--understood as for the mechanism but is not yet quantitatively conclusive.

\begin{figure}
\begin{center}
\includegraphics[width=12.0cm]{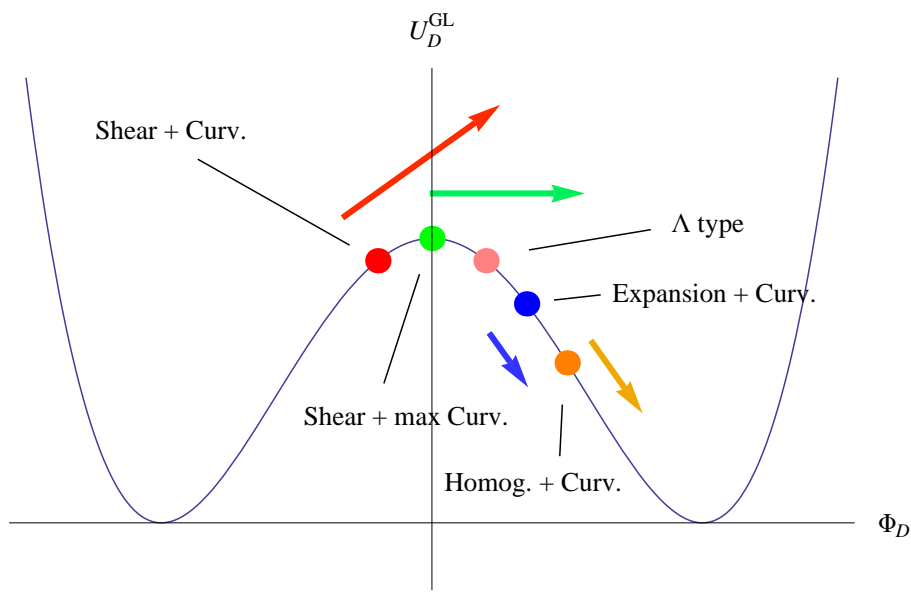}
\caption{The Ginzburg--Landau potential in arbitrary units and the possible 
initial conditions as well as their physical meaning.
All conditions possess some curvature $\CW_\initial\CD <0$.
The arrows schematically indicate the amplitude of the morphon's initial speed $\dot\Phi_\initial\CD$. 
In the order of the points (from left to right): the first two points dominated by shear fluctuations (red, green) are obtained for 
$\CQ_\initial\CD < 0 \Leftrightarrow \dot\Phi_\initial\CD^{2} > 2(H_\initial\CD^{2}+ k_\initial\CD )$; 
the next points dominated by expansion fluctuations (pink, blue) for $ \dot\Phi_\initial\CD^{2} < 2(H_\initial\CD^{2}+ k_\initial\CD )$,
where the de Sitter--$\Lambda$ equivalent case has a stiff morphon $\dot\Phi_\initial\CD = 0$; 
the homogeneous case (last point, orange) is obtained for  $ \dot\Phi_\initial\CD^{2} = 2(H_\initial\CD^{2}+ k_\initial\CD )$.\label{fig:GLpotential}}
\end{center}
\end{figure}

\begin{figure}
\begin{center}
\includegraphics[width=12.0cm]{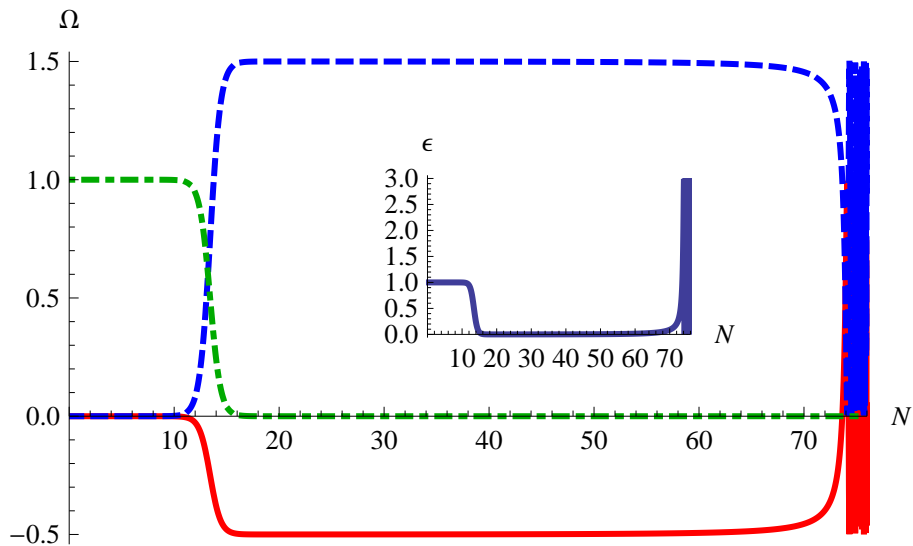}
\caption{The energy densities $\Omega^\CD_\CQ$ (solid, red), 
$\Omega^\CD_\CW$ (dashed, blue), 
and $\Omega^\CD_k$ (dotdashed, green)  
for one of the cases of Fig.~\ref{fig:GLpotential} as a function of the number of e--folds $N=\ln a_\CD$.
The initial value of the homogeneous part of the intrinsic curvature  
has been taken arbitrary large ({\it e.g.} $\Omega^\initial\CD_{k} \simeq 1$) 
to underline the fact that it vanishes anyway after a few e--folds. 
In the inset, the slow--roll parameter $\epsilon$ for the same configuration is shown.
Since $\epsilon =\Omega^\CD_k + {\dot\Phi}_\CD^2/2$ always holds, 
$\epsilon\simeq 1$ in the pre--inflation era when $\Omega^\CD_k\gg\Omega^\CD_\CQ,\Omega^\CD_\CW$,
and $\epsilon \simeq 0$ during inflation when $U_\CD^{GL}\gg {\dot\Phi}_\CD^2/2$.\label{fig:epsilon}
\label{fig:omega}}
\end{center}
\end{figure}

Once the minimum $\Phi_0$ is fixed,  the evolution of the morphon,
given the integrability condition (\ref{eq:integrability}), is practically
independent of the initial 
conditions\footnote{$U_{0}$ determines the inflation's onset time and influences also its duration.
Due to the necessity to recover the CMB temperature fluctuations, 
``classical" inflationary models suffer from fine--tuning issues, once perturbation theory is performed on the inflaton. 
Such is not the case here, since our model tackles the bulk effect of inflation within a highly non--linear regime,
without invoking perturbation theory. 
Hence, though a lack, the absence of perturbative constraints allows us to get rid of fine--tuning.
} \cite{Linde:1985ub}.
In Fig.~\ref{fig:GLpotential}
we show how all acceptable initial conditions 
are reinterpreted in terms of the curvature, and expansion/shear fluctuations.
For any of these types we find the behavior shown in the inset of Fig.~\ref{fig:omega}:
instead of simple inflation $\epsilon<1$, the choice of $U_\CD^{GL}$ gives us
slow--roll inflation $\epsilon\ll 1$ during which the energy density budget is dominated
by the curvature term $\Omega^\CD_\CW = - \CW_\CD / 6 H_\CD^2 \simeq 3/2$,
and a negative kinematical backreaction density 
$\Omega^\CD_\CQ = - \CQ_\CD / 6 H_\CD^2 \simeq -1/2$, i.e. by expansion variance, 
whatever the initial value of the homogeneous part of the curvature, 
$\Omega^\initial\CD_k= - k_\initial\CD / a_\initial\CD^2H_\initial\CD^2=1-\Omega^\initial\CD_\CQ-\Omega^\initial\CD_\CW$, is. 
At the end of the inflationary era, the morphon oscillates down its potential 
and, because of the order of the polynomial (\ref{GLpotential}), 
the effective scale factor behaves as if it were dominated by dust matter, $a_\CD \propto t^{2/3}$, 
and the effective deceleration parameter $q^{\CD} = - {\ddot a}_{\CD} (a_\CD H^2_\CD)^{-1} = 2 \Omega^\CD_\CQ$ oscillates around $q^\CD \cong 1/2$. 
While $\Omega^\CD_\CW$ oscillates around $3/4$ and $\Omega^\CD_\CQ$ around $1/4$, $\CW_\CD$ and $\CQ_\CD$ tend to vanish, 
i.e. our model generically tends to be -- on average -- homogeneous and quasi--isotropic.

Let us sum up the evolution scenario driven by kinematical backreaction $\CQ_\CD$:
thanks to a favorable configuration of the latter and of  the average curvature, 
a) a period of accelerated expansion kicks out ($\CQ_\CD$ rises more rapidly than $H_\CD^2$),
b) an inflationary era follows during which the expansion variance stays positive and almost constant ($\CQ_\CD \simeq 3 H_\CD^2 \simeq const.$),
c) the outcome is a globally inflated universe, hence exponentially smooth, where gravity and shear pull back, 
causing on time--average a negative kinematical backreaction $\bar \CQ_\CD \simeq -2 / 3 t^2$. 
Hence, the inhomogeneities wash out at the end of the process, so does the acceleration, hereby providing a natural graceful exit, 
univocally based, should we stress, on mere gravitation.

\section{Discussion} 

Looking at a portion of the classical vacuum within Einstein's theory, 
we obtained,
by foliating spacetime and spatially averaging the scalar parts of the Einstein
vacuum equations,
a Friedmann--like evolution of the scale factor, however,
driven by an effective scalar field. 
Starting with a sufficiently flat potential, 
the backreaction mechanism initiates a metamorphosis of the geometrical properties
of space that goes along with an exponential inflationary phase. 
This dynamics tends to flatten the averaged scalar curvature and suppresses fluctuations.
Although the present scenario just prescribes the initial conditions for a follow--up pre--heating, 
the same mechanism sets the conditions to provoke an intrinsic exit scenario. 
Note also that, while acceleration was associated to the formation of inhomogeneities in the Late Universe, {\it e.g.} \cite{Rasanen:2005zy},
the application of the same mechanism to the Early Universe tends to homogenize spatial hypersurfaces. 

Let us list some of the immediate consequences of this mechanism.
First, due to its classical nature, the inflaton's mass is no longer limited by the Planck/SUSY scale
(see {\it e.g.} \cite{McDonald:1999hd}).
Secondly, inflation is already possible for the unmodified Hilbert action, 
which of course does not exclude the need for improvements of Einstein's theory.
Third, inflation, often dubbed to be unsustainable in inhomogeneous spacetimes \cite{Goldwirth:1991rj}
(though see {\it e.g.} \cite{IguchiPerezBergliaffa}),
could occur despite the natural chaotic (inhomogeneous) initial conditions 
invoked by the theory \cite{Turok:2002yq}
(the reason for this behavior is essentially that 
we do not study the stability of the fluctuations on a homogeneous reference background
but that of a background--free general average).
 
Once this average correctly computed, one can address the issue of fluctuations
which one expects to be the seeds of large--scale structure and of the CMB 
spectrum; the absence of such a prediction is certainly a dearth that we should overcome (by formulating of a fluctuation theory about 
the average).
In a joined paper \cite{phasespace}, we improved the present model by 
proving that a foliation into flat space sections is unstable and is attracted by an inhomogeneous negative--curvature state, 
and also by proving that the ``empty" model we considered is a generic attractor of ``filled'' models under some conditions.
In a forthcoming paper, we shall address a more general model that includes matter, radiation and fundamental scalar field inhomogeneities. 
If sources are present, there are interesting interactions with a morphon field,
the latter being always present in the case of an inhomogeneous cosmology.
 
The presented scenario points to a huge potential of studying scalar field models in the Early Universe. 
We gave the simplest conceivable model that generates inflation out of the classical inhomogeneous vacuum.
Contrary to the quasi--Newtonian standard picture, 
we render the bulk effect of curvature responsible for inflation. 
In other words,
more than being acquitted of preventing inflation, 
inhomogeneities, when treated non--perturbatively, 
could be the actual cause of it.

\medskip
\noindent
{\footnotesize{\it Acknowledgments:}
Thanks go to  Alexandre 
Arbey, Bruce 
Bassett,  Henk 
van Elst, Syksy 
R\"{a}s\"{a}nen, Xavier 
Roy and Dominik 
Schwarz for valuable discussions, and to Xavier 
Roy
for checking the results.
This work is supported by ``F\'ed\'eration de Physique Andr\'e--Marie Amp\`ere'' of Universit\'e Lyon 1 and \'Ecole Normale Sup\'erieure de Lyon.}

\end{document}